\documentclass[aps,showpacs,superscriptaddress,nofootinbib,amsmath,amssymb]{revtex4-1}
\usepackage{graphicx}
\usepackage{dcolumn}
\usepackage{bm}
\usepackage{lipsum}
\usepackage{multirow}
\usepackage{adjustbox}
\graphicspath{{./eps/}}

\begin{document}

\title{Neutrino-nucleus cross sections in $^{12}C$ and $^{40}Ar$ with KDAR neutrinos}

\def\a              {\alpha}
\def\b              {\beta}
\def\m              {\mu}
\def\n              {\nu}
\def\ti             {\tilde}
\def\e              {\varepsilon}
\def\r              {\rho}
\def\s              {\sigma}
\def\g              {\gamma}
\def\d              {\delta}
\def\t              {\tau}

\author{F. \surname{Akbar}}

\author{M. Sajjad \surname{Athar}}
\email{sajathar@gmail.com}

\author{S. K. \surname{Singh}}
\affiliation{Department of Physics, Aligarh Muslim University, Aligarh-202002, India}

\begin{abstract} 
High intensity monoenergetic muon neutrinos of energy 236 MeV from kaon decay at rest (KDAR) at the medium energy proton accelerator facilities like J-PARC 
 and Fermilab are proposed to be used for making precision
measurements of neutrino-nucleus cross sections in $^{12}C$ and $^{40}Ar$ and perform neutrino oscillation experiments in $\nu_\mu \to \nu_\mu$ 
and $\nu_\mu \to \nu_e$ modes. In view of these developments, we study the theoretical uncertainties arising due to the nuclear 
 medium effects in the neutrino-nucleus cross sections as
well as in the angular and energy distributions of the charged
leptons produced in the charged current (CC) induced reactions by $\nu_\mu$ and $\nu_e$ in $^{12}C$  and $^{40}Ar$ in the energy region of
$E_{\nu_e(\nu_\mu)}<$ 300 MeV. The calculations have been done 
in a microscopic model using the local density approximation which takes into account the nuclear effects due to the Fermi motion,
binding energy and long range correlations. The results 
are compared with the other calculations available in the literature.
 \end{abstract}
\pacs{12.15.-y,13.15+g,24.10Cn,25.30Pt}
\maketitle
\section{Introduction}\label{Intro}
The two body leptonic decay mode of the charged kaon decay-at-rest (KDAR) i.e. $K^+ \to \mu^+ \nu_\mu$, B.R. 63.55$\pm$1.1$\%$~\cite{Olive:2016xmw} 
provides a unique and important source of monoenergetic 
muon neutrinos of energy 236 MeV. These neutrinos may be used to make high precision measurements of neutrino-nucleus cross sections 
for the charged current (CC) induced weak quasielastic (QE) 
production of muons from the various nuclear targets. The high precision neutrino-nucleus cross section measured with the well defined 
monoenergetic beam of muon neutrinos may serve as benchmark 
for validating many theoretical models currently being used to describe the nuclear medium effects in QE reactions~\cite{Katori:2016yel,Alvarez-Ruso:2014bla}
relevant for the analysis of present 
day neutrino experiments in the low energy region of a few hundred
MeVs~\cite{Abazajian:2012ys,Abe:2011sj,Antonello:2015lea,Katori:2011uq,Chen:2007ae,Dracos:pro61,Baussan:2013zcy,Vinning:2017izb,
Cao:2014bea,Adey:2015iha,Liu:2013zjn, Ajimura:2017fld,Harada:2016vlb,Harada:2015gla,Axani:2015dha,Spitz:2014hwa,Spitz:2012gp,grange}.

These KDAR neutrinos are proposed to be used as a probe to study the new neutrino oscillation modes to sterile neutrinos 
i.e. $\nu_\mu \to \nu_s$ by preforming the oscillation experiments in 
$\nu_\mu \to \nu_\mu$ disappearance mode and studying the CC interactions of $\nu_\mu$ with nuclei
and/or performing the oscillation experiments in $\nu_\mu \to \nu_e$ appearance mode and 
studying the CC interaction of $\nu_e$ with nuclei
~\cite{Ajimura:2017fld,Harada:2016vlb,Harada:2015gla,Axani:2015dha,Spitz:2014hwa,Spitz:2012gp,grange}.
In the $\nu_\mu \to \nu_\mu$ disappearance mode, $\nu_\mu$ from the three body $K \mu_3$ decays
of charged kaons i.e. $K^+ \to \mu^+ \pi^0 \nu_\mu$ having continuous energy spectrum with the end point energy of 215 MeV constitute the major
source of background while in 
the $\nu_\mu \to \nu_e$ appearance mode, $\nu_e$ from the $K e_3$ decay mode of charged kaons i.e. $K^+ \to e^+ \pi^0 \nu_e$, 
 having continuous energy spectrum with
end point energy of 228 MeV constitute the major source of background. The background in both the channels from the decay in flight (DIF)
neutrinos from pions, kaons and other 
mesons corresponds to higher energies. With sufficiently improved energy resolution for the detection of the 
final muon and the electron produced respectively in the CC weak interaction of $\nu_\mu$ and $\nu_e$ with matter, the background events can be well
separated in energy from the signal events for the oscillation experiments corresponding to
$E_{\nu_{\mu}(\nu_e)}$ = 236 MeV. Moreover, it has been recently suggested~\cite{Rott:2016mzs} that the observation of
CC induced QE events with 
the monoenergetic neutrinos can also provide 
information about the dark matter which annihilates in its interaction with the solar matter in the center of the Sun into
quark-antiquark pairs and produces the charged kaons through the hadronization process.
 The monoenergetic muon neutrinos produced by these charged kaons through the $K \mu_2$ decays can be identified by comparing the on-source and 
off-source event rates in the terrestrial detectors provided the background events for $E_\nu \sim$ 236 MeV are well under 
 control in the $\nu$-oscillation experiments 
proposed with the KDAR neutrinos. 

The feasibility of such experiments with high intensity KDAR neutrinos requires an accelerator facility capable of producing $K^+$ mesons with a very high yield.
The 3 GeV proton accelerator facility at the J-PARC MLF facility in Tokai,
Japan~\cite{Ajimura:2017fld,Harada:2016vlb,Harada:2015gla,Axani:2015dha,Spitz:2014hwa} 
and the 8 GeV proton accelerator facility at the BNB source facility at the Fermilab, 
USA~\cite{Spitz:2014hwa,Spitz:2012gp,grange}
have the sufficient energy and power to produce high intensity charged kaons through the primary and/or secondary interactions 
of protons with the nuclear targets which would be stopped in the surrounding material and their decay would give 
intense beam of $\nu_\mu$. At the J-PARC facility the neutrino oscillation experiments in the appearance mode 
i.e. $\nu_\mu \to \nu_e$ as well as in the disappearance mode i.e. 
 $\nu_\mu \to \nu_\mu$  have been proposed respectively, through the JSNS experiment by the
Japanese group~\cite{Ajimura:2017fld,Harada:2016vlb,Harada:2015gla}, and the KPipe experiment by the MIT-Columbia
 group~\cite{Axani:2015dha,Spitz:2014hwa,Spitz:2012gp} using the liquid scintillator detector with active detector mass of 17 tons and  684 tons, respectively.
 At the Fermilab facility a neutrino oscillation experiment in the appearance mode i.e.
$\nu_\mu \to \nu_e$ has been proposed with 2 kton LArTPC detector~\cite{Spitz:2012gp,grange}. 

One of the major sources of systematic errors in these experiments is due to the $\nu_\mu$ flux arising from the
uncertainty in the $K^+$ production yields in the proton-nucleus interaction predicted by the hadronic models for the kaon production and could be as large as 
75$\%$~\cite{Axani:2015dha,Spitz:2014hwa,Agostinelli:2002hh,Mokhov:2004aa}. The other 
source of systematic errors is due to the uncertainty in the $\nu_\mu(\nu_e)-$nucleus cross sections for $E_{\nu_{\mu}(\nu_e)}$ = 236 
MeV arising due to the nuclear medium effects~\cite{Katori:2016yel,Alvarez-Ruso:2014bla} and is the subject of the present work.

The present simulation studies~\cite{Axani:2015dha,Spitz:2014hwa,Spitz:2012gp}, for estimating the neutrino oscillation parameters, 
 use the neutrino nucleus cross sections for the KDAR neutrinos on  $^{12}$C and $^{40}$Ar as predicted by the 
NuWro generator~\cite{Juszczak:2009qa}
 which are reported to be 
 about 25$\%$ smaller than the predictions 
by the GENIE Monte Carlo generator~\cite{Andreopoulos:2009rq} and the results of Martini et al.~\cite{Martini:2011wp,Martini:2009uj}.
 In the low energy region, the short range correlations and the meson exchange currents(MEC) are not expected to play an 
  important role~\cite{Nieves:2012yz,Benhar:2005dj,Martini:2012uc}, but the effects of Pauli blocking, 
Fermi motion and the long range RPA correlations are found to be quite important. This has been shown by many theoretical 
 attempts~\cite{Kosmas:1996fh,Engel:1996zt,Auerbach:1997ay,Singh:1998md,Volpe:2000zn,Hayes:1999ew,Kolbe:1999au,Kolbe:2003ys,Nieves:2004wx,Paar:2007fi,Auerbach:2002tw,Paar:2008zza,Nieves:2017lij} made to explain the $\nu_\mu-^{12}$C
cross section measured in the LSND experiment~\cite{Albert:1994xs,Athanassopoulos:1997rm,Auerbach:2002iy} with the pion decay in flight (DIF) muon neutrinos in
the energy region of
$E_{\nu_\mu} <$ 320 MeV with $<E_{\nu_\mu}> =$ 150 MeV. These effects could therefore be very important in the energy region of KDAR neutrinos. 

In view of the recent interest in the proposed neutrino oscillation experiments in $\nu_\mu \to \nu_\mu$ and $\nu_\mu \to \nu_e$
mode with liquid scintillator(LS) and LArTPC detectors
and the search of sterile neutrinos through the $\nu_\mu \to \nu_s$ mode; it is topical to study the uncertainties 
in the $\nu_\mu(\nu_e)-$nucleus cross sections in the low energy region relevant for the monoenergetic KDAR neutrinos. In this paper, we have studied 
the uncertainties in the neutrino-nucleus cross sections for the QE processes induced by the weak charged current
 interaction in $\nu_\mu(\nu_e)$ scattering from $^{12}C$ and $^{40}Ar$
nuclei relevant for the KDAR neutrinos with $E_{\nu_\mu} \le$ 300 MeV in a nuclear model using the local density
approximation
 which takes into account the effects of 
nuclear medium arising due to the Pauli Blocking, Fermi motion and the long range RPA correlations. The model has been used by us earlier to calculate quite 
satisfactorily the low energy neutrino cross sections relevant for the supernova, Michel and pion decay in flight(DIF) neutrino spectra
~\cite{SajjadAthar:2004yf,Athar:2005ca,SajjadAthar:2005ke,Chauhan:2017tgf}.
 We report the results on the energy dependence of the total cross section $\sigma(E_\nu)$ for $E_\nu < $300MeV, and the angular distributions
($\frac{d\sigma}{dcos\theta_l}$) and the kinetic energy distributions ($\frac{d\sigma}{dT_l}$) for the electron and the muon produced in the
CCQE reactions induced by $\nu_e$ and $\nu_\mu$ at $E_{\nu}=$ 236 MeV
in $^{12}$C and $^{40}$Ar and compare these results with the other theoretical calculations available in the literature.

\begin{figure}
\begin{center}
\includegraphics[height=.2\textheight,width=0.25\textwidth]{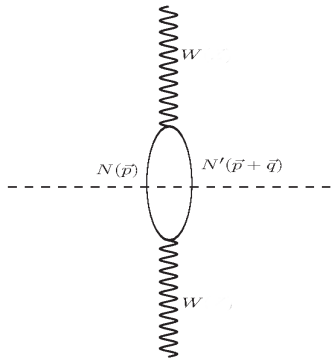}
\caption{Diagrammatic representation of the particle - hole(p-h) excitation induced by $W$ boson in the large mass limit of intermediate 
 vector boson($M_W \rightarrow \infty$).}
\label{fig:neutrinoselfenergy}
\end{center}
\end{figure}
\begin{figure}
\begin{center}
\includegraphics[height=.2\textheight,width=0.60\textwidth]{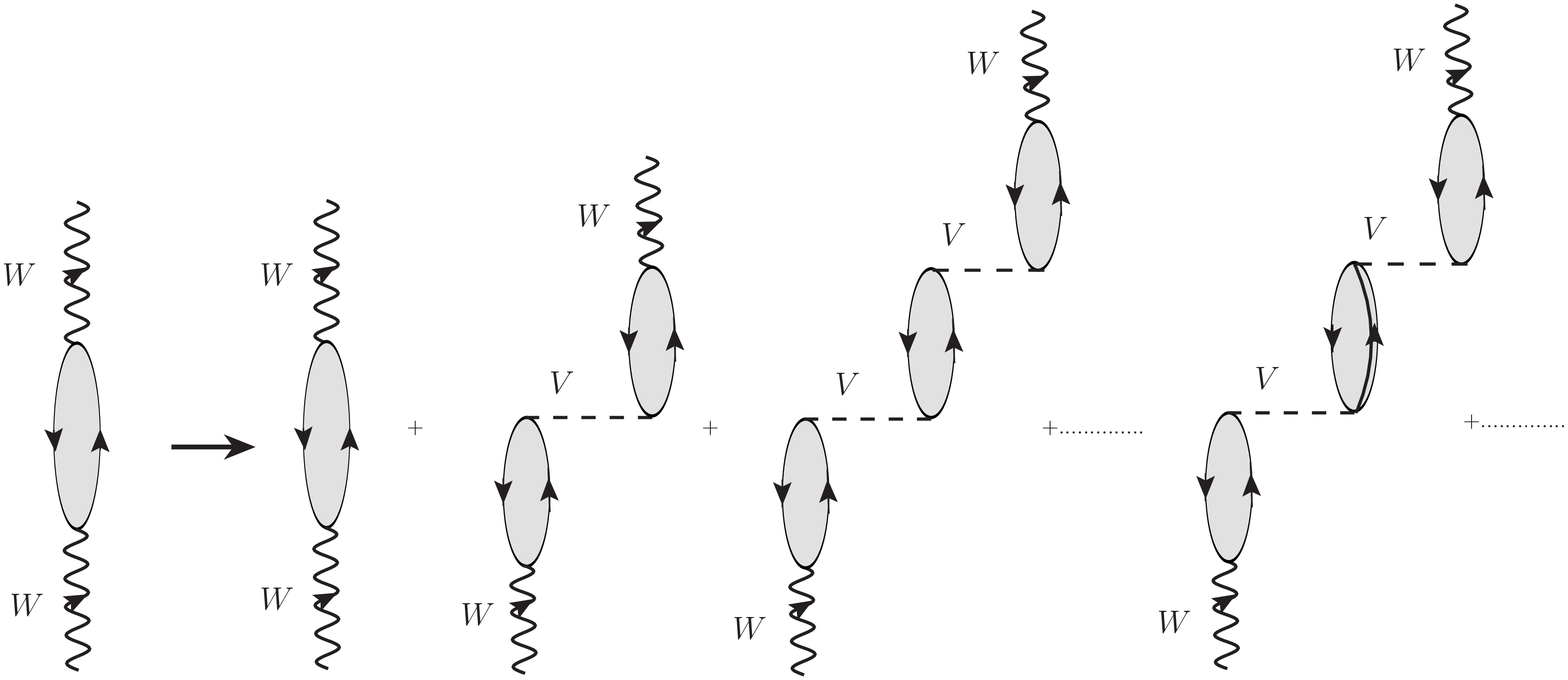}
\caption{RPA effects in the 1p1h contribution to the W self energy, where particle-hole, $\Delta$-hole, $\Delta$-$\Delta$, etc. excitations 
contribute.}
\label{fg:fig2}
\end{center}
\end{figure}
\section{Formalism}\label{Formalism}
The reaction for the CC neutrino interaction with a nucleus is given by 
\begin{equation}
 \nu_l + ^{A}_{Z}\!X \to l^- + _{Z+1}^{A}\!Y~~~~~(l=e,\mu)
\end{equation}
for which the basic process is 
\begin{equation}\label{rxn1}
 \nu_l(k) + n(p) \to l^-(k') + p(p').
\end{equation}
$^{A}_{Z}\!X(_{Z+1}^{A}\!Y)$ is the initial(final) nucleus, and $k,~k'$ are the four momenta of the incoming and outgoing lepton and $p,~p^\prime$ are the 
 four momenta of the initial and final
nucleon, respectively. The invariant matrix element given in Eq.(\ref{rxn1}) is written as
\begin{eqnarray}\label{qe_lep_matrix}
{\cal M}=\frac{G_F}{\sqrt{2}}\cos\theta_c~l_\mu~J^\mu
\end{eqnarray}
where $G_F$ is the Fermi coupling constant (=1.16639$\times 10^{-5}~GeV^{-2}$), 
$\theta_c(=13.1^0)$ is the Cabibbo angle. The leptonic weak current is 
given by
\begin{eqnarray}\label{lep_curr}
l_\mu&=&\bar{u}(k^\prime)\gamma_\mu(1 - \gamma_5)u(k),
\end{eqnarray}
$J^\mu$ is the hadronic current given by
\begin{eqnarray}\label{had_curr}
J^\mu&=&\bar{u}(p')\Gamma^\mu u(p),
\end{eqnarray}
with
\begin{eqnarray}\label{eq:had_int}
\Gamma^\mu=F_1^V(Q^2)\gamma^\mu+F_2^V(Q^2)i\sigma^{\mu\nu}\frac{q_\nu}{2M} 
+ F_A(Q^2)\gamma^\mu\gamma^5 
+ F_P(Q^2) \frac{q^\mu}{M}\gamma^5  ,
\end{eqnarray}
 $Q^2(=-q^2)~\geq 0$ is the four momentum transfer square and $M$ is the nucleon mass. $F_{1,2}^V(Q^2)$
 are the isovector vector form factors and $F_A(Q^2)$, $F_P(Q^2)$ are the axial and pseudoscalar form factors, respectively.
 We have not considered the contribution from the second class currents.
\begin{figure}
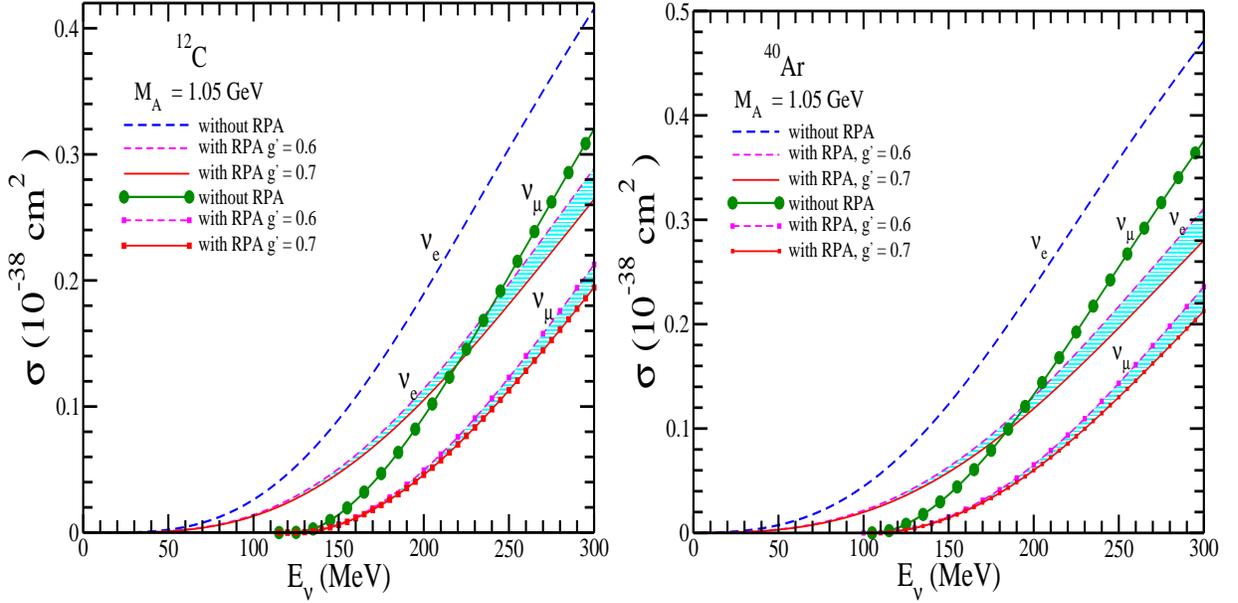

 \includegraphics[width=8cm,height=8cm] {sigma_C_combined_Gp_band.eps}
  \includegraphics[width=8cm,height=8cm]{sigma_Ar_combined_Gp_band.eps}
 \caption{$\sigma$  vs $E_{\nu_l}$, for $\nu_l$($l$ = $e^-,~\mu^-$) induced scattering on $^{12}C$(left panel) and $^{40}Ar$(right panel) nuclear targets. 
 The dashed line(line with circles) represents $\nu_e$($\nu_\mu$) cross section obtained in the LFGM without RPA effects, while the bands upper(lower) 
 represents $\nu_e$($\nu_\mu$) cross section with
 RPA. The bands correspond to the variation of $g^\prime$ in the range of 0.6-0.7.}
 \label{sigmacband}
\end{figure}

\begin{figure}
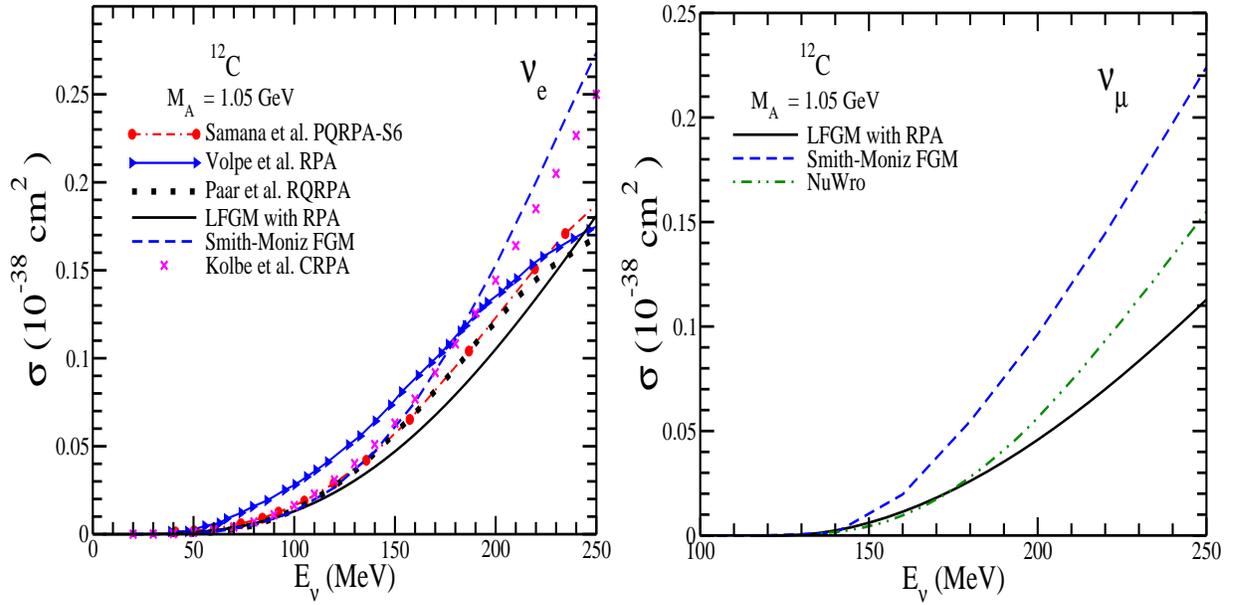

 \includegraphics[width=8cm,height=8cm]{sigma_C_e-1.eps}
 \includegraphics[width=8cm,height=8cm]{sigma_C_mu.eps}\\
 \caption{$\sigma$  vs $E_{\nu_l}$, for $\nu_e$(left panel) and $\nu_\mu$(right panel) CCQE scattering cross sections on $^{12}C$ in the 
 LFGM model with RPA effect(solid line), the results of NuWro event generator taken from Ref.~\cite{Spitz:2012gp} (dashed double-dotted line),
 Volpe et al.~\cite{Volpe:2000zn} in RPA (triangle right), Kolbe et al.~\cite{Kolbe:1999au} in CRPA~(cross),
 Paar et al.~\cite{Paar:2007fi} in RQRPA~(dotted line),
 Samana et al.~\cite{Samana:2011zz} in PQRPA (circle), and 
 Smith and Moniz~\cite{Smith:1972xh} in RFGM (dashed line).}
\label{sigmac}
\end{figure}

\begin{figure}
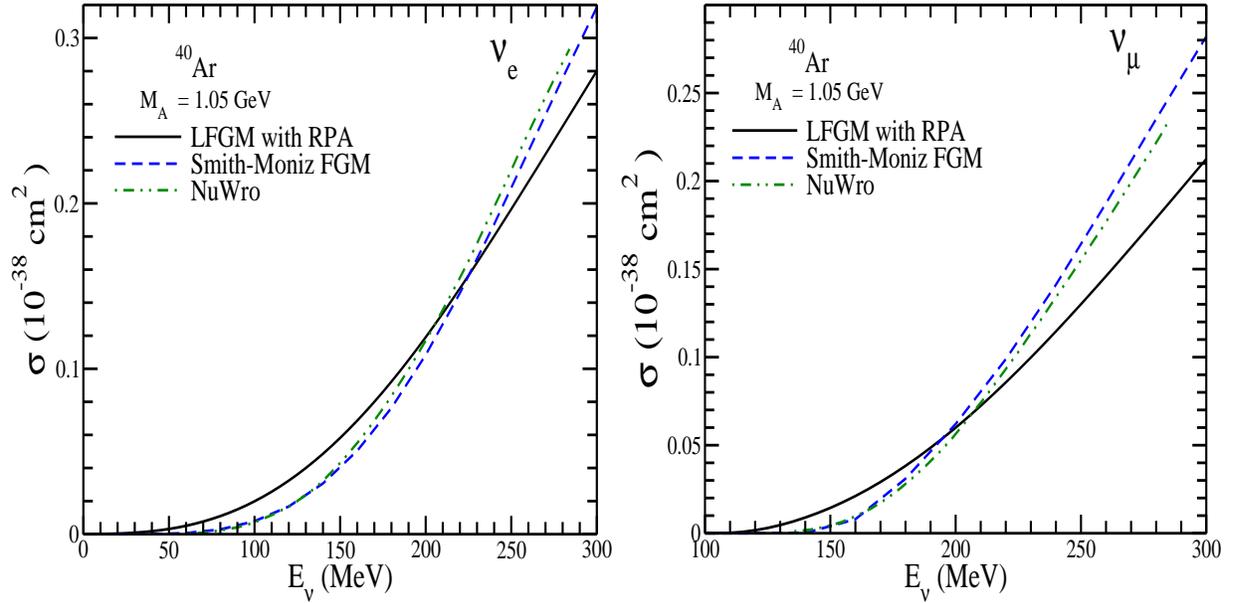

 \includegraphics[width=8cm,height=8cm]{sigma_Ar_e.eps}
 \includegraphics[width=8cm,height=8cm]{sigma_Ar_mu.eps}\\
 \caption{$\sigma$  vs $E_{\nu_l}$, for $\nu_e$(left panel) and $\nu_\mu$(right panel) on $^{40}Ar$ in the LFGM model with RPA effect(solid line), 
   the results of NuWro generator taken from Ref.~\cite{Spitz:2012gp} (dashed - double dotted), and 
   Smith and Moniz in RFGM~\cite{Smith:1972xh,Andreopoulos:2009rq} (dashed line).}
\label{sigmaar}
\end{figure}

The hadronic current contains isovector vector form factors $F_{1,2}^V(Q^2)$ of the nucleons, which are given as
\begin{equation}\label{f1v_f2v}
F_{1,2}^V(Q^2)=F_{1,2}^p(Q^2)- F_{1,2}^n(Q^2) 
\end{equation}
where $F_{1}^{p(n)}(Q^2)$ and $F_{2}^{p(n)}(Q^2)$ are the Dirac and Pauli form factors of proton(neutron) 
which in turn are expressed in terms of the
experimentally determined Sach's electric $G_E^{p,n}(Q^2)$ and magnetic $G_M^{p,n}(Q^2)$ form factors as 
\begin{eqnarray}\label{f1pn_f2pn}
F_1^{p,n}(Q^2)&=&\left(1+\frac{Q^2}{4M^2}\right)^{-1}~\left[G_E^{p,n}(Q^2)+\frac{Q^2}{4M^2}~G_M^{p,n}(Q^2)\right]\\
F_2^{p,n}(Q^2)&=&\left(1+\frac{Q^2}{4M^2}\right)^{-1}~\left[G_M^{p,n}(Q^2)-G_E^{p,n}(Q^2)\right]
\end{eqnarray}
$G_E^{p,n}(Q^2)$ and $G_M^{p,n}(Q^2)$ are the electric and magnetic Sachs form factors and for the numerical calculations we have 
 used the parameterization of Bradford et al.~\cite{Bradford:2006yz}.

 The isovector axial form factor is obtained from the quasielastic neutrino and antineutrino scattering 
as well as from the pion electroproduction data and is parameterized as
\begin{equation}\label{fa}
F_A(Q^2)=F_A(0)~\left[1+\frac{Q^2}{M_A^2}\right]^{-2};~~F_A(0)=-1.267.
\end{equation}

The pseudoscalar form factor $F_P(Q^2)$ is dominated by the pion pole and is given in terms of the axial vector form factor $F_A(Q^2)$
using the Goldberger-Treiman(GT) relation~\cite{LlewellynSmith:1971zm}:
\begin{equation}\label{fp}
F_P(Q^2)=\frac{2M^2F_A(Q^2)}{m_\pi^2+Q^2}.
\end{equation}

The differential cross section corresponding to Eq.~\ref{rxn1} is given by
\begin{equation}\label{sig0}
\sigma_{0}({\bf q^2, k^\prime, p}) =  \frac{1}{4\pi} \frac{k^2}{E_{\nu} E_{l}} \frac{M^2}{E_{n} E_{p}} \bar\Sigma\Sigma |{\cal M}^{2}| \delta(q_0 + 
E_{n}  - E_{p}),
\end{equation}
where $q_0=E_{\nu_l}-E_l$, $E_n=\sqrt{{|\bf p|}^2 + M_n^2}$ and $E_p=\sqrt{{|{\bf p + q}|}^2 + M_p^2}$. The matrix element square is obtained by using
Eq.(\ref{qe_lep_matrix}) and is given by
\begin{equation}\label{mat_quasi}
{|{\cal M}|^2}=\frac{G_F^2}{2}~{ L}_{\mu\nu} {J}^{\mu\nu}.
\end{equation}
In Eq.(\ref{mat_quasi}), ${ L}_{\mu\nu}$ is the leptonic tensor calculated to be
\begin{eqnarray}\label{lep_tens}
{L}_{\mu\nu}&=&{\bar\Sigma}\Sigma{l_\mu}^\dagger l_\nu=L_{\mu\nu}^{S} - i L_{\mu\nu}^{A},~~~~\mbox{with}\\
L_{\mu\nu}^{S}&=&8~\left[k_\mu k_\nu^\prime+k_\mu^\prime k_\nu-g_{\mu\nu}~k\cdot k^\prime\right]~~~~\mbox{and}\nonumber\\
L_{\mu\nu}^{A}&=&8~\epsilon_{\mu\nu\alpha\beta}~k^{\prime \alpha} k^{\beta},
\end{eqnarray}

The hadronic tensor ${J}^{\mu\nu}$ given by:
\begin{eqnarray}\label{had_tens}
J^{\mu\nu}&=&\bar{\Sigma}\Sigma J^{\mu\dagger} J^\nu,
\end{eqnarray}
where $J^{\mu}$ defined in Eq.(\ref{had_curr}) with Eqs.(\ref{f1v_f2v}), (\ref{fa}) and (\ref{fp}) has been used for the numerical calculations. 
The detailed expression for the hadronic tensor $J^{\mu\nu}$ is given in Ref.~\cite{Akbar:2015yda}.

When the processes given by Eq.~(\ref{rxn1}) take place in a nucleus, various nuclear medium effects like 
Pauli blocking, Fermi motion, binding energy corrections and nucleon correlations, etc. come into play. Moreover, the charged lepton produced 
in the final state moves in the Coulomb field of the residual nucleus and which affects its energy and momenta. We have taken into account these 
effects which are briefly discussed below:
 
 \begin{enumerate}
\item In the standard treatment of the Fermi Gas Model applied to neutrino reactions 
 the quantum states of the nucleons inside the nucleus are filled up to a Fermi momentum $p_{F}$, given by 
$p_{F}= \left[3 \pi^2 \rho\right]^{\frac{1}{3}}$, where $\rho$ is the density of the nucleus. In a nuclear reaction, the momentum of the initial 
nucleon  $p$ is therefore constrained to be 
 $p < p_{F}$ and $p^\prime (= |{\bf p} + {\bf q}|) > p_{F}^\prime$, where $p_{F}$ 
 is the Fermi momentum of the initial nucleon target in the Fermi sea, and 
 $p_{F}^\prime$ is the Fermi momentum of the outgoing nucleon.
 The total energies of the initial($i$) and final($f$) nucleons are $E_i=\sqrt{{|\bf p|}^2+M_i^2}$ and $E_f=\sqrt{|{\bf p} + {\bf q}|^2+M_f^2}$.
 In this model the Fermi momentum and energy are constrained to be determined by the nuclear density which is constant.  
 
  In the local Fermi gas model(LFGM), the Fermi momenta of the initial and final nucleons are not constant but depend on the interaction point 
$\vec r$ and are given by $p_{F_n}(r)$ and $p_{F_p}(r)$ for neutron and proton, respectively, where 
 $p_{F_n}(r)= \left[3 \pi^2 \rho_n (r)\right]^{\frac{1}{3}}$ and $p_{F_p}(r)= \left[3\pi^2 \rho_p (r)\right]^{\frac{1}{3}}$,
$\rho_n (r)$ and $\rho_p (r)$ being the neutron and proton nuclear densities, respectively.
 We use the proton density $\rho_p (r) = \frac{Z}{A}\rho (r)$ and neutron density given by $\rho_n (r) = \frac{A-Z}{A}\rho (r)$, where $\rho (r)$  is 
 determined
 experimentally by the electron-nucleus scattering experiments~\cite{Vries:1974}. We use modified harmonic oscillator(MHO) density 
 \begin{eqnarray}\label{MHO}
 \rho(r) = \rho(0) \left[ 1 + a \left( \frac{r}{R} \right)^{2} exp\left[ -\left( \frac{r}{R} \right)^{2} \right] \right]
 \end{eqnarray}
for $^{12}C$  and 2-parameter Fermi density(2pF) 
\begin{eqnarray}\label{2pF}
 \rho(r) = \frac{\rho(0)} {\left[ 1 +  exp\left( \frac{r-R}{a} \right)\right]}
 \end{eqnarray}
for $^{40}Ar$ with $R$ and $a$ as the density parameters and the parameters are taken from the Refs.~\cite{Vries:1974, GarciaRecio:1991wk}. In Table-\ref{tab:Q-value_BE}, we show the nuclear density and other parameters needed 
for the numerical calculations in this paper. 
 
  In the local density approximation(LDA), the cross section($\sigma$) for the $\nu_l$ scattering from a nucleon moving in the 
  nucleus with a momentum ${\bf p}$ is given by~\cite{Singh:1993rg}:
\begin{equation}\label{sigma_lda}
\sigma(q^2, k^\prime) = \int 2 d{\bf r} d{\bf p} \frac{1}{(2\pi)^3} n_n({\bf p(r)}) [1-n_{p}({\bf p(r)} + {\bf q(r)})] \sigma_{0}({\bf q^2, k^\prime, p}), 
\end{equation}
 where $\sigma_{0}$ is given by Eq.\ref{sig0}. In the above expression, $n_n({\bf p(r)})$ and $n_{p}({\bf p(r)} + {\bf q(r)})$ represent the occupation numbers 
 for the neutron and proton respectively i.e. at a given position ${\bf r}$, $n_n({\bf p(r)})$=1 for $p \le p_{F_n}(r)$, and 0
 otherwise, and $n_{p}({\bf p(r)} + {\bf q(r)})$=1 for $|{\bf p(r)} + {\bf q(r)}| \ge p_{F_p}(r)$, and 0 otherwise.
  
Instead of using Eqs.~\ref{sig0} and~\ref{sigma_lda}, we use the methods of many body field theory~\cite{Fetter}, where the reaction cross section for the process $\nu_l + n 
\to l^- + p$ in a nuclear medium is given in terms of the imaginary part of the Lindhard function $U_N(q_0,\vec q)$ corresponding to the p-h 
 excitation diagram shown in Fig.\ref{fig:neutrinoselfenergy}~\cite{Singh:1993rg}.
 This imaginary part $U_N(q_0,\vec q)$ is obtained by cutting the $W$ self energy diagram along the horizontal line(Fig.
 \ref{fig:neutrinoselfenergy}) and applying the
Cutkowsky rules~\cite{Itzykson}. This is equivalent to replacing the expression
 \begin{equation}
\int \frac{d\bf{p}}{(2\pi)^3}{n_n(\bf{p})} [1-n_{p}({\bf p} + {\bf q})] \frac{M_n M_p}{E_{n}({\bf p}) E_{p}(\bf p+\bf  q)}\delta[q_0+E_n-E_p]
\end{equation}
occurring in Eq.(\ref{sigma_lda}) by $-(1/{\pi})$Im$ U_N(q_0,\vec{q})$, where 
\begin{equation}
U_N(q_0,{\bf q}) = \int \frac{d{\bf p}}{(2\pi)^3} \frac{M_{n}M_{p}}{E_{n}({\bf p})E_{p}(\bf p + \bf q)} \frac{n_{n}({\bf p})
[1-n_{p}({\bf p} + {\bf q})]}{q_{0}+E_{n}({\bf p})-E_{p}(\bf p+\bf q)+i\epsilon }.
\end{equation}
  The imaginary part of the Lindhard function is calculated to be~\cite{Singh:1993rg}:
\begin{equation}\label{lind}
Im ~ U_N (q_{0}, {\bf q}) = -\frac{1}{2\pi} \frac{M_{p} M_{n}}{|{\bf q}|} [E_{F_{1}} - A]
\end{equation}
 for $q^2 < 0, E_{F_2} - q_{0} < E_{F_{1}}$ and $\frac{-q_0 + |{\bf q}|\sqrt{1 - \frac{4 M^2}{q^2} }}{2} < E_{F_{1}}$, 
 
 otherwise $Im ~U_N = 0$.

 In the above expression $E_{F_{1(2)}} = \sqrt{p_{F_{n(p)}}^2 + M_{n(p)}^{2}}~$, and

$A = Max \left[ M_{n}, E_{F_{2}} - q_{0}, \frac{-q_0 + |{\bf q}|\sqrt{1 - \frac{4 M^2}{q^2} }}{2}\right]$.
 
 \item When the reaction $\nu_l + n \rightarrow l^- + p$ takes place in the nucleus, the first consideration is the Q value which inhibits
 the reaction in the nucleus. The experimental Q values corresponding to the g.s. $\rightarrow$ g.s. transition are given in Table-I
 for the two nuclei.  We also introduce $Q_F(r)= E_{F_{2}}(r) -  E_{F_{1}}(r)$ to take into account the difference in the Fermi levels of the initial and final nuclei,
 which  results in an effective value of $Q = Q-Q_F(r)$ to be used in the local Fermi Gas model.
 These considerations imply that $q_0$ should be modified to $q_0^{\text eff}(r) = q_0-(Q-Q_F(r))$ in the calculation of the Lindhard function in Eq.\ref{lind}.
 
 \item In the charged current reaction, the energy and momentum  of  the outgoing charged  lepton are modified due 
to the Coulomb interaction with the final nucleus. The Coulomb distortion effect on the outgoing lepton 
  has been taken into account in an effective momentum approximation(EMA)~\cite{Engel:1997fy,Kim:1996ua,Traini:2001kz,Aste:2005wc}
 in which the lepton momentum and energy are modified by replacing $E_l$ by $E_l+V_c(r)$. The form of the Coulomb potential $V_{c}(r)$ considered here is:
\begin{equation}\label{cool}
V_{c}(r) =-\alpha~ 4\pi\left(\frac{1}{r}\int_{0}^{r}\frac{\rho_{p}(r^\prime)}{Z}r^{\prime 2}dr^{\prime} + \int_{r}^{\infty}\frac{\rho_{p}
(r^\prime)}{Z}r^{\prime}dr^{\prime}\right),
\end{equation}
where $\alpha$ is fine  structure constant and $\rho_{p}(r)$ is the proton density of the final nucleus. 

Incorporation of these considerations results in the modification of the argument of the Lindhard function (Eq.\ref{lind}), i.e.
\[Im U_N (q_0, {\bf q})~\longrightarrow~Im U_N (q_0^{\text eff}(r)~-~V_c(r), {\bf q}).\]

 With the inclusion of these nuclear effects, the cross section $\sigma(E_\nu)$ is written as
 {\footnotesize
\begin{eqnarray}\label{xsection_medeffect}
\sigma(E_\nu)=-2{G_F}^2\cos^2{\theta_c}\int^{r_{max}}_{r_{min}} r^2 dr \int^{{k^\prime}_{max}}_{{k^\prime}_{min}}k^\prime dk^\prime
 \int_{Q^{2}_{min}}^{Q^{2}_{max}}dQ^{2}\frac{1}{E_{\nu_l}^{2} E_l} L_{\mu\nu}J^{\mu\nu}Im U_N (q_0^{\text eff}(r) - V_c(r), {\bf q}).~~~~
\end{eqnarray}}
 We must point out that in the above expression the outgoing lepton momentum and energy are $r$-dependent i.e. $k^\prime=k^\prime(r)$ and 
 $E_l=E_l(r)$, and only in the asymptotic limit ($r \rightarrow \infty$) they become independent of r. With the incorporation of
 the Coulomb effect, $E_l(r)$ is modified to $E_l(r) + V_{c}(r)$, and $|\vec k^\prime(r)|=\sqrt{(E_l(r) + V_{c}(r))^2 - m_l^2}$. 
 Accordingly the energy transfer $q_0$ modifies to $q_0^{\text eff}(r) = q_0^{\text eff}(r) - V_c(r)$, and the three momentum transfer $\vec q$ modifies 
  to $\vec q(r)= \vec k - \vec k^\prime(r)$.
\begin{table}\label{tab:Q-value_BE}
\begin{tabular}{|c|c|c|c|c|c|}\hline\hline
Nucleus &  Q-Value($\nu$)& $R_p$ & $R_n$ & a \\
        &       (MeV)      & (fm)&(fm)&$(fm)^\ast$\\ \hline
$^{12}C$& 17.84  &1.69 &1.692  &1.082(MHO)\\
$^{40}Ar$& 3.64  & 3.47& 3.64 & 0.569(2pF)\\ \hline
\end{tabular}
\caption{Fermi momentum and Q-value of the reaction for $^{12}C$ and $^{40}Ar$ nuclear targets. Last three columns are the parameters for MHO and 2pF densities~\cite{Nieves:2004wx, Vries:1974, GarciaRecio:1991wk}.
 $^{\ast}$ is dimensionless for the MHO density.}
 \end{table}
 
 \item In the nucleus the strength of the electroweak couplings may change from their free nucleon values due to the presence of 
strongly interacting nucleons. Conservation of vector current (CVC) forbids any change in the charge coupling while the magnetic 
and the axial vector couplings are likely to change from their free nucleon values. There exists considerable work in understanding the 
 quenching of magnetic moment and axial charge in nuclei due to nucleon-nucleon correlations.
 In our approach these are reflected in the modification  of nuclear response in longitudinal and transverse channels leading to some reduction. 
 We calculate this reduction in the vector-axial(VA) and axial-axial(AA) response functions due to the long range nucleon-nucleon correlations treated in the 
 random phase approximation(RPA), which has been diagrammatically shown in Fig.(\ref{fg:fig2}). 

The weak nucleon current described by Eq.(\ref{had_curr}) gives in the non-relativistic limit, terms like $F_A \vec{\sigma}\tau_+$ 
and $i F_2^V \frac{\vec{\sigma}\times \vec{q}}{2M}\tau_+$ which generate spin-isospin transitions in nuclei. While the term 
$i F_2^V \frac{\vec{\sigma}\times \vec{q}}{2M}\tau_+$ couples with the transverse excitations, the term  $F_A \vec{\sigma}\tau_+$ 
couples with the transverse as well as longitudinal channels. These channels produce different RPA responses in the longitudinal and 
transverse channels due to the different NN potential in these channels when the diagrams of Fig.(\ref{fg:fig2}) are summed up. 
As a consequence a term proportional to $F^2_A \delta_{ij}$ in $J^{ij}$ is replaced by $J^{ij}_{RPA}$ as \cite{Akbar:2015yda}:
\begin{equation}\label{f2a_rpa}
J^{ij}\rightarrow J^{ij}_{RPA}= F^2_A{Im U_N}\left[\frac{{\bf{\hat{q_i}}{\hat{q_j}}}}{1-U_NV_l}+\frac{\delta_{ij}-{\bf{\hat{q_i}}{\hat{q_j}}}}
{1-U_NV_t}\right],
\end{equation}

where the first and second terms show the modification in $J^{ij}$ in longitudinal and transverse channels. 
In Eq.(\ref{f2a_rpa}), $V_l$ and $V_t$ are the longitudinal and transverse parts of the nucleon-nucleon potential calculated using $\pi$ and $\rho$ 
exchanges and are given by 
\begin{eqnarray}\label{longi_part}
V_l(q) = \frac{f^2}{m_\pi^2}\left[\frac{q^2}{-q^2+m_\pi^2}{\left(\frac{\Lambda_\pi^2-m_\pi^2}{\Lambda_\pi^2-q^2}\right)^2}+g^\prime\right],\nonumber\\
V_t(q) = \frac{f^2}{m_\pi^2}\left[\frac{q^2}{-q^2+m^2_\rho}{C_\rho}{\left(\frac{{\Lambda_\rho}^2-m^2_\rho}{{\Lambda_\rho}^2-q^2}\right)^2}+g^\prime
\right],\end{eqnarray}
 where $\frac{f^{2}}{4\pi}$ = 0.8, $\Lambda_\pi$ = 1.3 GeV, $C_\rho$ = 2, $\Lambda_\rho$ = 2.5 GeV, $m_\pi$ and $m_\rho$ are the pion and rho meson masses, 
and $g^\prime$ is the Landau-Migdal parameter taken to be $0.7$ 
which has been used quite successfully to explain weak processes in nuclei
~\cite{Singh:1993rg,Singh:1998md,SajjadAthar:2005ke,Chauhan:2017tgf}. 
 However, in some recent works, the Valencia group~\cite{Nieves:2004wx,Nieves:2017lij,Valverde:2006zn} has used the value of $g'=$ 0.63.
 We have, therefore, studied the dependence of 
$g'$ on the total cross section as well as lepton energy and angular distributions by varying $g'$ in the range of 0.6 to 0.7.

The effect of the $\Delta$ degrees of freedom in the nuclear medium is included in the calculation of the RPA response by considering 
the effect of ph-$\Delta$h and $\Delta$h-$\Delta$h excitations. 
This is done by replacing $U_N \rightarrow U_N^{\prime}=U_N+U_\Delta$, where $U_\Delta$ is the Lindhard function for the
$\Delta$h excitation in the nuclear medium. The expressions for $U_N$ and $U_\Delta$ are taken from Ref.\cite{Oset1}. 
The different couplings of $N$ and $\Delta$ are incorporated in $U_N$ and $U_\Delta$ and then the same interaction strengths 
($V_l$ and $V_t$) are used to calculate the RPA response. 

With the incorporation of these nuclear medium effects the expression for the total scattering cross section $\sigma(E_\nu)$ is given 
by Eq.(\ref{xsection_medeffect}) with $J^{\mu \nu}$ replaced by $J^{\mu \nu}_{RPA}$(defined in Eq. (\ref{f2a_rpa})) i.e.
\begin{eqnarray}\label{xsection_final}
\sigma(E_\nu)&=&-2{G_F}^2 cos^{2}\theta_c \int^{r_{max}}_{r_{min}} r^2 dr \int^{{k^\prime}_{max}}_{{k^\prime}_{min}}k^\prime dk^\prime
 \int_{Q^{2}_{min}}^{Q^{2}_{max}}dQ^{2}\frac{1}{E_{\nu_l}^{2} E_l}  L_{\mu\nu}J^{\mu \nu}_{RPA} Im{U_N}(q_0^{\text eff}(r)~-~V_c(r)),~~~~~~
\end{eqnarray}
where $J^{\mu \nu}_{RPA}$ is the hadronic tensor with its various components modified due to long range correlation effects treated in RPA as it is 
shown in Eq.(\ref{f2a_rpa}) for the leading term proportional to $F_A^2$. The explicit expressions for $J^{\mu \nu}_{RPA}$ is given in the Ref.\cite{Akbar:2015yda}.
\end{enumerate}

\begin{figure}
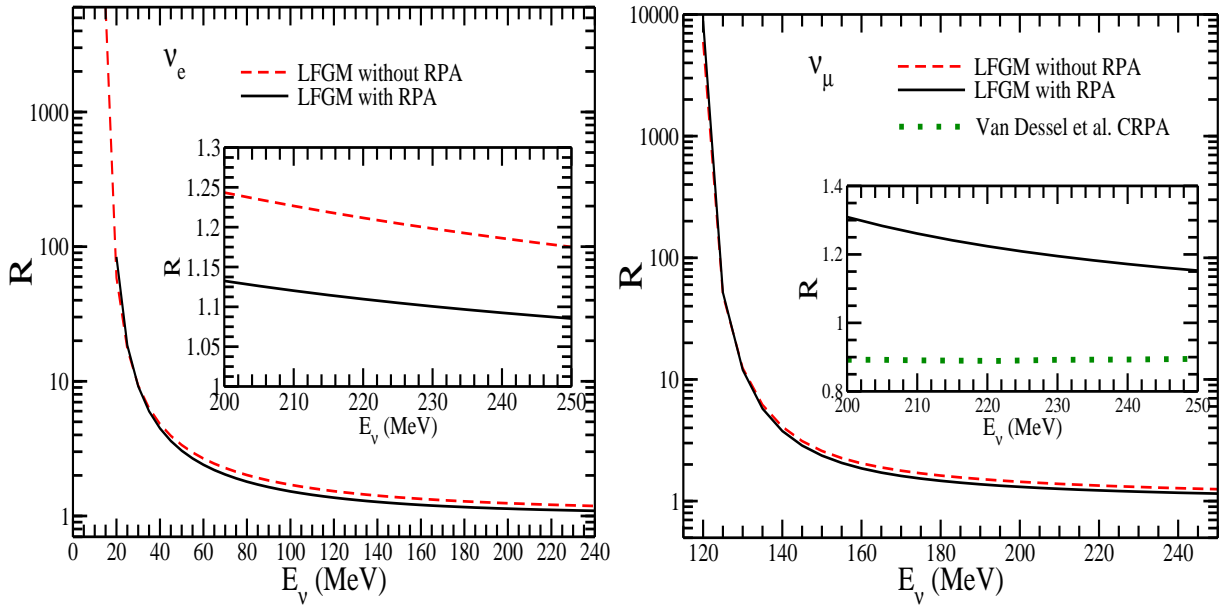

 \includegraphics[width=8cm,height=8cm]{ratio-sigma-nue.eps}
  \includegraphics[width=8cm,height=8cm]{ratio-sigma-numu.eps}
  \caption{Ratio, $R=\frac{\sigma^{^{40}Ar}_{\nu_{_l}}}{\sigma^{^{12}C}_{\nu_{_l}}}$ vs $E_{\nu}$ for $\nu_e$(left panel) and $\nu_\mu$(right panel). Solid(dashed)
  line represents 
  the results obtained using LFGM with(without) RPA. In the inset of $\nu_e$ case(left panel), the results of the ratio are obtained without(dashed line) 
  and with(solid line) RPA.
  In the inset of $\nu_\mu$ case(right panel), the results are compared with the results of 
  Van Dessel et al.~\cite{VanDessel:2017ery} in CRPA(dotted line).}\label{ratio_sigma}
\end{figure}

\begin{figure}
 \includegraphics[width=12cm,height=8cm]{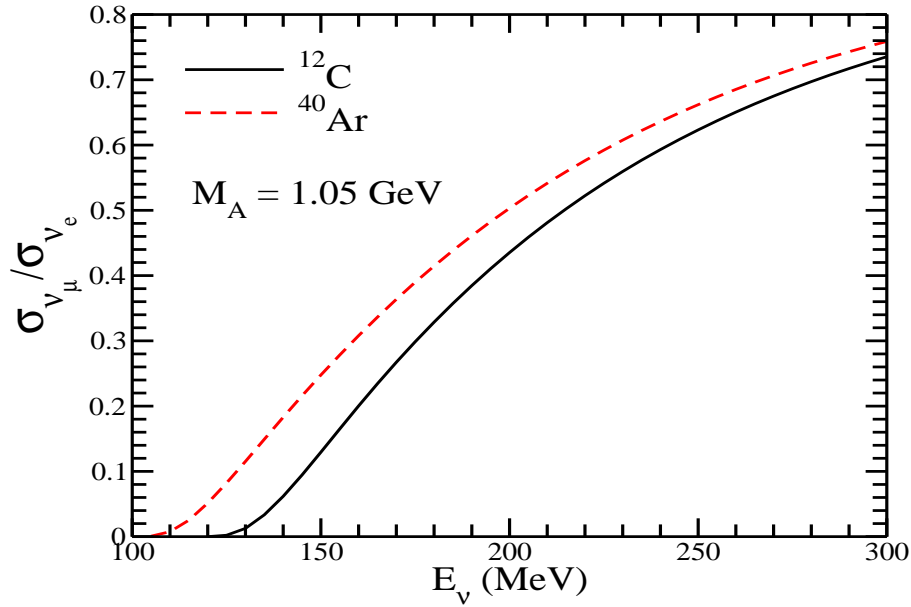}
 \caption{Ratio of $\nu_\mu$ to $\nu_e$ scattering cross sections $\frac{\sigma_{\nu_\mu}}{\sigma_{\nu_e}}$  vs $E_\nu$ for $^{12}C$(solid line) and 
 $^{40}Ar$(dashed line) in the LFGM with RPA.}
\label{ratioemu}
\end{figure}

\begin{figure}
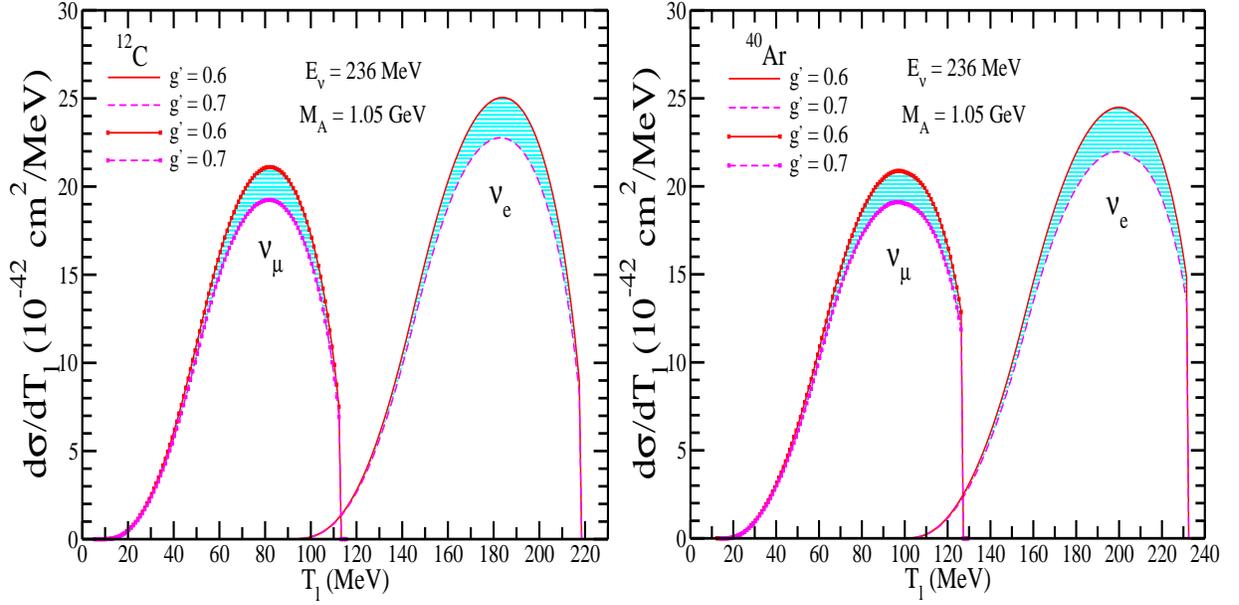

 \includegraphics[height=8cm,width=8cm]{dsdT_C_wRPA_Gp_band.eps}
 \includegraphics[height=8cm,width=8cm]{dsdT_Ar_wRPA_Gp_0.6.eps}
 \caption{$\frac{d\sigma}{dT_l}$ vs $T_l$ for $\nu_l$ induced processes on $^{12}C$(left panel) and $^{40}Ar$(right panel) nuclear targets at $E_\nu = $236 MeV.
 The results are obtained using
 LFGM with RPA. The variation of $g'$ from 0.6 to 0.7 is represented by the band. The curves on the left(right) side of each panel represent the results for $\mu^-(e^-)$ kinetic energy 
 distribution induced by $\nu_\mu(\nu_e)$ scattering.}
 \label{dTmugp}
\end{figure}

\begin{figure}
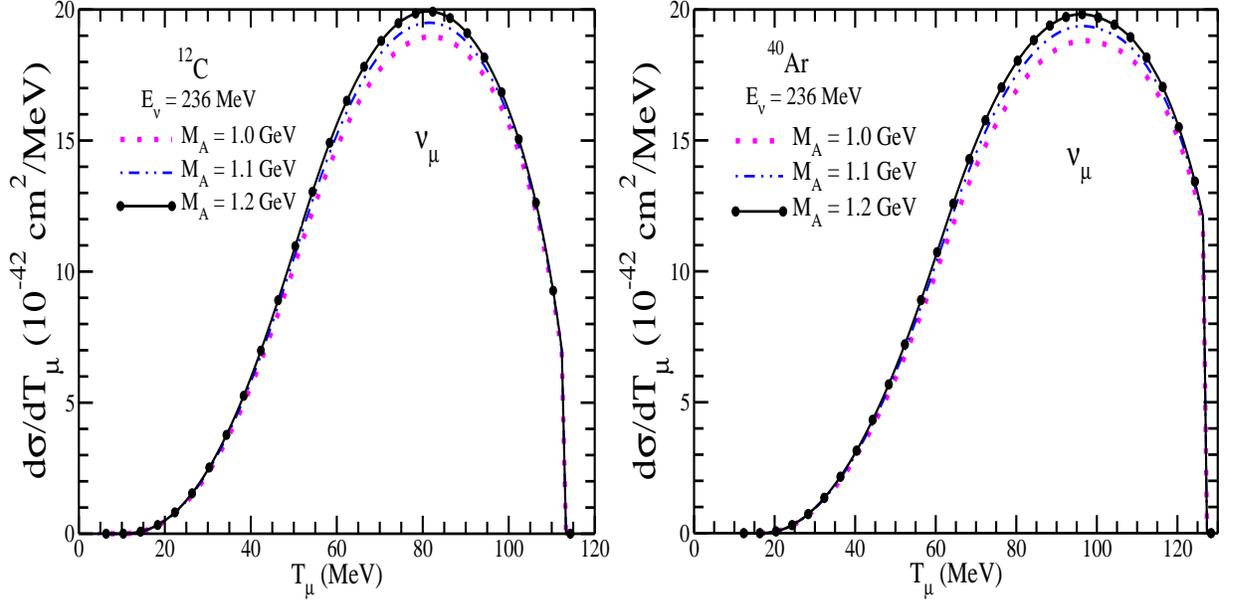

 \includegraphics[height=8cm,width=8cm]{lepton_mass_dep_dTmu_C12.eps}
 \includegraphics[height=8cm,width=8cm]{dsdT_Ar_wRPA_Ma.eps}
 \caption{ $\frac{d\sigma}{dT_\mu}$ vs $T_\mu$ for $\nu_\mu$ induced processes on $^{12}C$(left panel) and $^{40}Ar$(right panel) nuclear targets at $E_\nu = $236 MeV.
 The results are obtained using LFGM with RPA for the different values of $M_A$ viz. $M_A =$ 1.0 GeV(dotted line), 1.1 GeV(dash double-dotted line) 
 and 1.2 GeV(circle), respectively.}\label{dTmu}
\end{figure}

\begin{figure}
 \includegraphics[height=8cm,width=12cm]{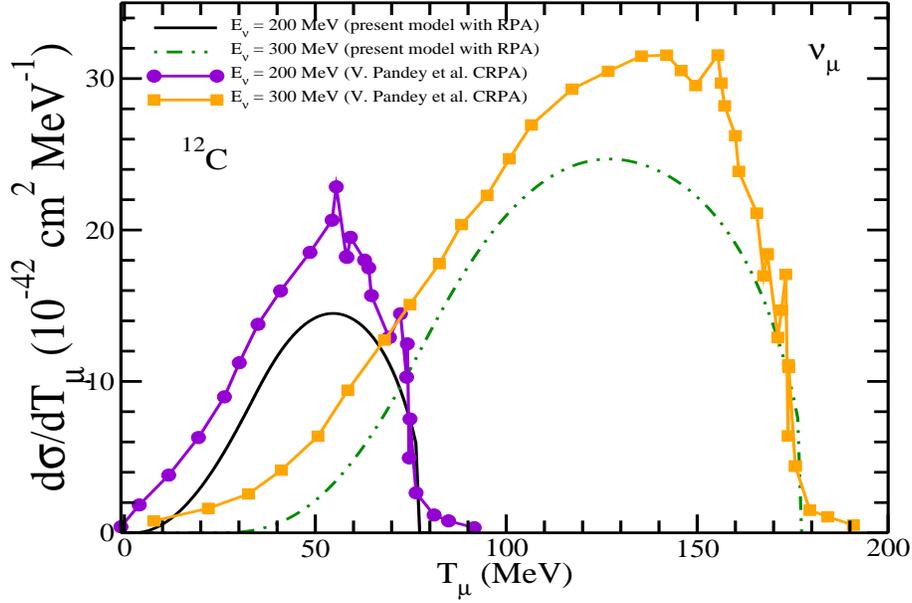}
\caption{Kinetic energy distribution, $\frac{d\sigma}{dT_\mu}$ vs $T_\mu$ for $\nu_\mu$ induced process on $^{12}C$ nuclear target.
The results of LFGM with RPA are represented 
by solid line at $E_\nu =$ 200 MeV and by dash double-dotted line at $E_\nu =$ 300 MeV. The results obtained from the Ref.~\cite{Pandey:2014tza}
at $E_\nu =$ 200 MeV(circle) and $E_\nu =$ 300 MeV(square) are also presented in the figure.}\label{pandey}
\end{figure}

\begin{figure}
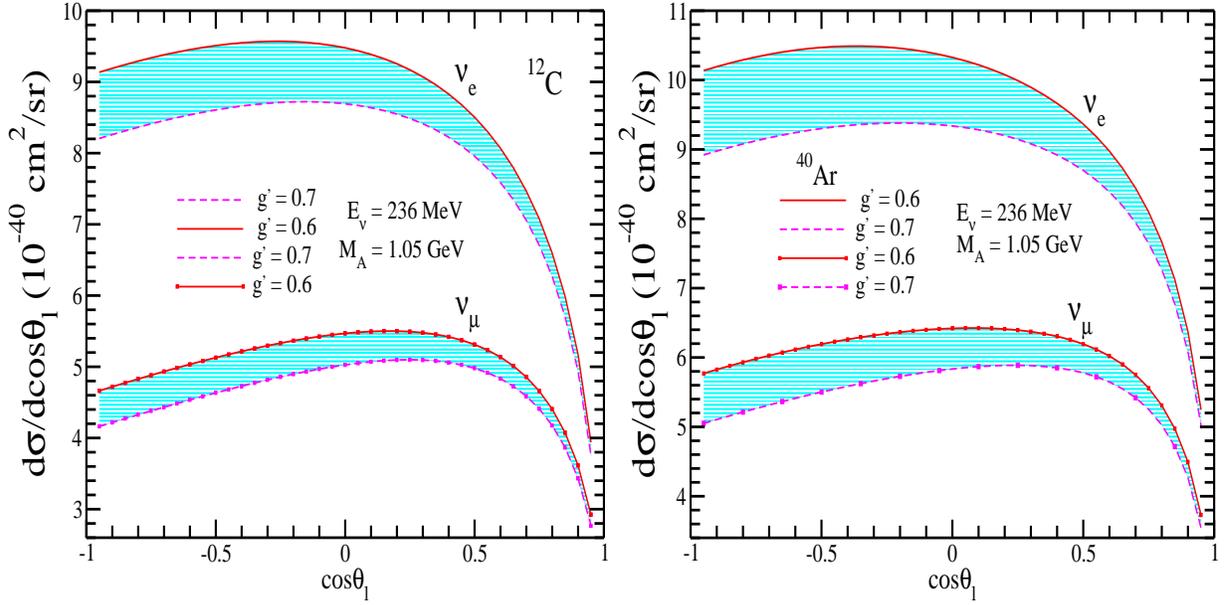

 \includegraphics[width=8cm,height=8cm]{dsdcos_C_wRPA_Gp_0.6.eps}
  \includegraphics[width=8cm,height=8cm]{dsdcos_Ar_wRPA_Gp_0.6.eps}
  \caption{$\frac{d\sigma}{cos{\theta_l}}$ vs $cos{\theta_l}$ for $\nu_l$ induced process on $^{12}C$(left panel) and $^{40}Ar$(right panel) at $E_{\nu}$=236MeV,
 obtained by using LFGM with RPA.
 The variation of $g'$ from 0.6 to 0.7 is represented by the band. Solid and dashed lines present the results of $\nu_e$ scattering and dotted and 
 dash double-dotted lines 
  represent the results of $\nu_\mu$ scattering.}\label{dcos}
\end{figure}

\begin{figure}
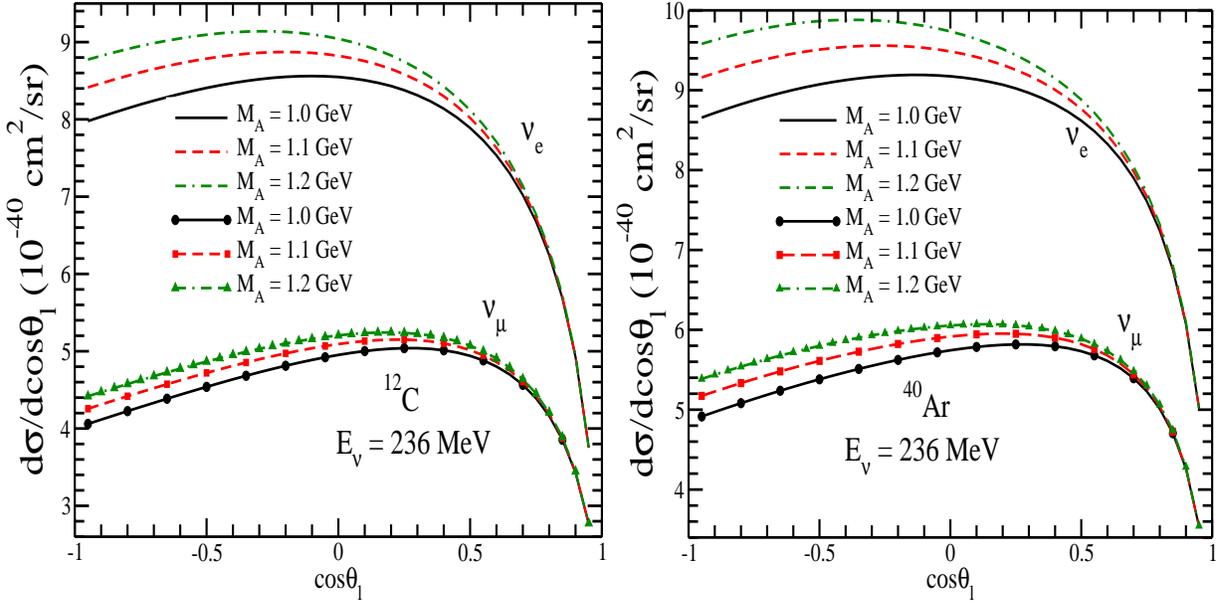

 \includegraphics[height=8cm,width=8cm]{lepton_mass_dep_dcosth_C12.eps}
 \includegraphics[height=8cm,width=8cm]{lepton_mass_dep_dcosth_Ar40.eps}
 \caption{$\frac{d\sigma}{cos{\theta_l}}$ vs $cos{\theta_l}$ for $\nu_e$(top curves) and $\nu_\mu$(bottom curves)
 induced processes on $^{12}C$(left panel) and $^{40}Ar$(right panel) 
 nuclear targets at $E_\nu = $236 MeV, obtained by using LFGM with RPA for the different values of $M_A$ viz. $M_A =$ 1.0 GeV(circle), 
 1.1 GeV(dash double-dotted line) and 1.2 GeV(dotted line).}
 \label{dcos_Ma}
\end{figure}

\section{Results and discussion}\label{Results}

For the numerical calculations, we have used Eq.~\ref{xsection_medeffect} to obtain the results for the charged current
$\nu_e$ and $\nu_\mu$ scattering cross sections on the nuclear targets 
in the local Fermi gas model(LFGM) with the inclusion of Fermi momentum and Pauli blocking,
and Eq.~\ref{xsection_final} when RPA effects are also included. Furthermore, we have taken 
 Coulomb distortion effect on the outgoing charged lepton in both cases using EMA with the Coulomb potential given in Eq.~\ref{cool}. 

In Fig.~\ref{sigmacband}, we present the results of $\nu_l$(l=$e,\mu$) induced charged lepton production cross sections $\sigma$ vs $E_{\nu_l}$ 
 in $^{12}C$ and $^{40}Ar$, respectively. We find a large reduction in the cross section due to the nuclear medium effects. For example, in the case of 
  $\nu_e$ scattering on $^{12}C$($^{40}Ar$) nuclear 
  targets, when the cross section is obtained using the LFGM without RPA effects, the reduction in the cross section from the free nucleon case(not shown here)
  is $\sim$50\%(35\%) at $E_{\nu_e}=$ 150 MeV, $\sim$38$\%$(20$\%$) at $E_{\nu_e}=$ 200 MeV
and $\sim$30$\%$(15$\%$) at $E_{\nu_e}=$ 236 MeV. When the RPA effects are also taken into account there is a further reduction in the cross section which is about
$\sim$48\%(53\%) at $E_{\nu_e}=$ 150 MeV, $\sim$45$\%$(50$\%$) at $E_{\nu_e}=$ 200 MeV
and $\sim$42$\%$(47$\%$) at $E_{\nu_e}=$ 236 MeV. In the case of $\nu_\mu$ scattering, this reduction is $\sim$85$\%$(65$\%$) at $E_{\nu_\mu}=$ 150 MeV, $\sim$60$\%$(43$\%$) at $E_{\nu_\mu}=$ 200 MeV
and $\sim$47$\%$(30$\%$) at $E_{\nu_\mu}=$ 236 MeV without the RPA correlation and a further reduction of  
$\sim$55$\%$(60$\%$) at $E_{\nu_\mu}=$ 150 MeV, $\sim$50$\%$(55$\%$) at $E_{\nu_\mu}=$ 200 MeV
and $\sim$45$\%$(50$\%$) at $E_{\nu_\mu}=$ 236 MeV, when RPA effects are included. We have also shown in these figures, the dependence of the cross section on 
$g'$, the Landau-Migdal parameter, used in Eq.~(\ref{longi_part}) by varying the value of $g'$ in the range 0.6-0.7. 
The bands shown in the figures correspond to the change in the cross section due to the variation of $g'$ in this range.
We find that with  $g'=0.7$ 
the cross section in $^{12}C$/$^{40}Ar$
  decreases by about 10\% for $\nu_l(l=e,\mu)$ scattering at 236 MeV from the results obtained with $g'=0.6$.
  
 In Figs.~\ref{sigmac} and \ref{sigmaar}, we have compared the present results in $\nu_e - ^{12}C$ with the results of 
 NuWro generator~\cite{Juszczak:2009qa}, Volpe et al.~\cite{Volpe:2000zn} in Random Phase Approximation, 
 Kolbe et al.~\cite{Kolbe:1999au} in Continuum Random Phase Approximation, Paar et al.~\cite{Paar:2007fi} in Relativistic Quasiparticle Random Phase Approximation, 
 Samana et al.~\cite{Samana:2011zz} in Projected Quasiparticle Random Phase Approximation-S6, and 
 Smith and Moniz~\cite{Smith:1972xh} in the relativistic Fermi gas model.
 In the case of $\nu_\mu - ^{12}C$, $\nu_e - ^{40}Ar$ and $\nu_\mu - ^{40}Ar$ induced processes, the cross section results are compared with the results of Smith and Moniz~\cite{Smith:1972xh} and 
 NuWro generator~\cite{Spitz:2012gp,Juszczak:2009qa} in which the nucleon spectral function of Benhar et al.~\cite{Benhar:2005dj} has been used.

In the preliminary simulation studies for determining the neutrino oscillation parameters Spitz et al.~\cite{Spitz:2014hwa,Spitz:2012gp} have used 
 NuWro~\cite{Juszczak:2009qa}
prediction of 1.3$\times$10$^{-38}$ cm$^2$ per neutron for the total cross section for $\nu_\mu-^{12}C$ and $\nu_\mu-^{40}Ar$ scattering at $E_\nu=$ 236 MeV.
 and the same value has also been used by Axani et al.~\cite{Axani:2015dha} for the $\nu_\mu-^{12}C$ cross section.
 In view of this we have studied the ratio R = $\frac{\sigma^{^{40}Ar}}{\sigma^{^{12}C}}$ as a function of $E_\nu$.
 In Fig.~\ref{ratio_sigma}, we have shown the results for R obtained using the present model with and without RPA effects
 for the $\nu_e$ and $\nu_\mu$ induced processes and also made a comparison with the recent results reported by Van Dessel et al.~\cite{VanDessel:2017ery} in CRPA
 for the $\nu_\mu$ induced process.
 
 The measurement of neutrino-nucleus cross section induced by 
 $\nu_\mu$  and $\nu_e$  and the 
 $\nu_\mu/\nu_e$ cross section ratio 
 is an important quantity in the analysis of $\nu_\mu \to \nu_e$ oscillations in the appearance channel. This ratio also provides an experimental validation 
 of the theoretical calculations of the various effects 
 arising due to the lepton mass dependent terms in the standard model specially the pseudoscalar form factors and the second
 class currents~\cite{Akbar:2015yda,Day:2012gb} 
 and could provide possible evidence of muon--electron non-universality.
 Moreover, this ratio is also a key parameter in improving the sensitivity of measuring the CP violation phase $\delta_{CP}$ in the future 
 experiments on neutrino oscillations~\cite{Huber:2007em}.
 We have plotted in Fig.~\ref{ratioemu}, the ratio of $\nu_\mu/\nu_e$ cross section as a function of energy in the low energy region relevant for the experiments
  with KDAR neutrinos.

In Fig.\ref{dTmugp}, we have presented the results of $\frac{d\sigma}{dT_l}$ vs $T_l$ ($l=e,\mu$), for $\nu_e$ and  $\nu_\mu$  induced processes in $^{12}C$(left panel) 
 and $^{40}Ar$(right panel) nuclear targets at $E_\nu = $236 MeV. The results are shown by varying $g'$ in the range 0.6 to 0.7 using $M_A$=1.05GeV.
 
In Fig.~\ref{dTmu}, we have presented the results for $\frac{d\sigma}{dT_\mu}$ vs $T_\mu$ in $^{12}C$ and $^{40}Ar$ 
nuclear targets at $E_\nu =$ 236 MeV by varying $M_A$ in the range of 1 - 1.2GeV. 
  We observe that there is very little sensitivity on $M_A$ in the case of kinetic energy distribution which is
 around 5\% when we vary the value of $M_A$ by 20\%.

In Fig.~\ref{pandey}, we show the energy distribution $\frac{d\sigma}{dT_\mu}$ vs $T_\mu$ at the two different neutrino energies, viz., $E_{\nu_\mu}=$ 200 MeV 
 and $E_{\nu_\mu}=$ 300 MeV on $^{12}C$ calculated in the present model and compare them with the results of Pandey et al.~\cite{Pandey:2014tza}.
  We see that there is wide variation in the energy distribution of muons predicted in these two models. 
  
In Fig.~\ref{dcos}, we have presented the results of the angular distribution $\frac{d\sigma}{d\cos\theta_l}$ vs $\cos\theta_l$ for $\nu_e$ and 
 $\nu_\mu$  induced processes
in $^{12}C$(left panel) and $^{40}Ar$(right panel) nuclear targets at $E_\nu = $236 MeV.
The results are obtained by varying $g'$ in the range 0.6 to 0.7 using $M_A$=1.05GeV.

In Fig.~\ref{dcos_Ma}, we have presented the results for $\nu_e$ and $\nu_\mu$ induced processes in $^{12}C$ and $^{40}Ar$ 
nuclear targets at $E_\nu =$ 236 MeV by varying $M_A$ in the range of 1 - 1.2GeV. 
  We observe that there is some sensitivity on $M_A$ in the case of angular distribution specially at the backward angles corresponding to higher $Q^2$.

\section{Summary and Conclusions}
We have presented a theoretical description of the inclusive quasielastic scattering for $\nu_e$ and $\nu_\mu$ scattering induced by the weak charged
 current on $^{12}C$ and $^{40}Ar$
relevant for the future experiments planned to be done using KDAR neutrinos. These KDAR neutrinos are monoenergetic with $E_{\nu_\mu}$=236MeV.
 The neutrino oscillation experiments in the $\nu_\mu \rightarrow \nu_\mu$ and $\nu_\mu \rightarrow \nu_e$ channels have 
 background from the KDAR neutrinos from the $K_{\mu_3}$ and $K_{e_3}$ decay modes with the 
 continuous energy spectrum of the muon neutrinos with $E_{\nu_\mu} \le$ 215MeV and the electron neutrinos with $E_{\nu_e} \le$ 235MeV.
 We have therefore studied the nuclear medium effects in the neutrino-nucleus cross sections in the energy region of $E_{\nu_l} \le 300MeV$ for QE
 scattering of $\nu_e$ and $\nu_\mu$ from 
 $^{12}C$ and $^{40}Ar$ nuclear targets proposed to be used in the future experiments planned at the JPARC and Fermilab facilities with liquid 
 scintillator(LS) and LArTPC detectors.
 
 The calculations have been done in a microscopic model of nucleus which takes into account the effect of the Fermi motion, binding energy and long range 
 nucleon-nucleon correlations through RPA. The method has been earlier applied successfully to reproduce the low energy neutrino-nucleus cross sections 
 observed in LSND, KARMEN and LAMPF experiments. The effect of Pauli
  blocking and RPA correlations is to drastically reduce the cross sections over the free nucleon case. In the energy region of $E_\nu=150-250MeV$,
  the overall reduction due to Pauli blocking and RPA 
  correlations in  $^{12}C$($^{40}Ar$) varies from the range of 70$\%$(75$\%$) to 55$\%$(53$\%$) in the case of 
  $\nu_e$-nucleus scattering and 95$\%$(85$\%$) to 68$\%$(63$\%$) in the case of 
  $\nu_\mu$-nucleus scattering. There is an uncertainty of about 10$\%$ due to the Landau-Migdal parameter used in the treatment of RPA correlations. The results 
  have been compared with the results of the other calculations in the literature. The cross section obtained using the present model with 
  RPA effect is around 50$\%$ smaller than the results of the Relativistic Fermi gas model(RFGM) of Smith and Moniz~\cite{Smith:1972xh}. The different treatment of 
  nucleon-nucleon 
  correlations in the various approaches discussed in section-\ref{Results} results in an uncertainty of about $25\%$ in the cross sections at $E_{\nu_\mu}$=236MeV.
  
  We have also presented the results for the energy and angular distributions for $e^-$ and $\mu^-$ produced in these reactions for a fixed neutrino energy of 
  $E_\nu$=236MeV. A comparison with the recent results of Pandey et al.~\cite{Pandey:2014tza} shows that the differences
  in the prediction of these two models are significant in the case of the energy distributions of the muons. 
  An experimental observation of the total cross sections and the differential cross sections in the forthcoming experiments 
  will be able to discriminate between various models of treating the nuclear medium effects in the neutrino-nucleus scattering at low energies.

 \end{document}